\documentclass[aps,pra,superscriptaddress,twocolumn,floatfix]{revtex4-1}
\usepackage{amsmath,amssymb}
\usepackage{bm}
\usepackage{graphicx}

\usepackage[colorlinks]{hyperref}
 
\begin{document}

\title{Macroscopic random Paschen-Back effect in ultracold atomic
gases}
\author{M. Modugno}
\affiliation{\mbox{Department of Theoretical Physics and History of Science, University of the
Basque Country UPV/EHU, 48080 Bilbao, Spain}}
\affiliation{IKERBASQUE Basque Foundation for Science, Bilbao, Spain}
\author{E. Ya. Sherman}
\affiliation{Department of Physical Chemistry, University of the Basque Country UPV/EHU,
48080 Bilbao, Spain}
\affiliation{IKERBASQUE Basque Foundation for Science, Bilbao, Spain}
\author{V. V. Konotop}
\affiliation{Centro de F\'isica Te\'orica e Computacional and Departamento de F\'isica, 
Faculdade de Ci\^encias, Universidade de Lisboa, 
Campo Grande 2, Edif\'icio C8, Lisboa 1749-016, Portugal}
\date{\today}

\begin{abstract}

We consider spin- and density-related properties of single-particle states 
in a one-dimensional system with random spin-orbit coupling. 
We show that the presence of an additional Zeeman field $\Delta$ induces both 
nonlinear spin polarization and delocalization of states localized at $\Delta=0$, corresponding to 
a random macroscopic analogue of the Paschen-Back effect. While the conventional Paschen-Back effect corresponds
to a saturated $\Delta-$dependence of the spin polarization, here the gradual suppression of the spin-orbit 
coupling effects by the Zeeman field is responsible both for the spin saturation and delocalization of the particles.
\end{abstract}

\maketitle

\section{Introduction}

Spin and mass dynamics caused by spin-orbit coupling (SOC) constitute one of the most important and interesting topics in modern 
solid-state and condensed-matter physics \cite{Dyakonov08,spielman2013,zhaih2012}. Recent experiments
with ultracold atomic gases have greatly extended the frontiers of this field, 
by realizing tunable artificial SOC, as well Zeeman fields, for Bose-Einstein condensates \cite{EXspielman2011} 
and Fermi gases \cite{wang2012,cheuk2012}. 
The possibility of studying the effects of strong SOC both
experimentally and theoretically has revealed a rich phenomenology of these systems 
(see e.g. \cite{EXjin2012,EXqu2013,stringari2012,zhang2012,lu2013}).
In one-dimensional settings this phenomenology has been enhanced by
the presence of additional potentials, such as lattices \cite{OL,OL_ZL,ZL} or artificial defects in the SOC ~\cite{LocSO}. 

A key topic in low-dimensional solid state \cite{Anderson,Berezinskii}
and cold atomic \cite{Flach1,Pikovsky,Larcher1,Aleiner} systems
is the localization of particles by disorder.
In the presence of SOC, the
localization was studied in Ref. \cite{Zhou} and in a 
quasi-periodic potential in Ref. \cite{LYKKC}, where 
a mobility edge was observed. 
Short-term spin and density dynamics were considered in Ref. \cite{mardonov2015}. 
Yet another type of SOC - the random one - is naturally present in solids 
\cite{Glazov2010,Glazov2011,Bindel2016}. It can also be designed in cold 
atomic matter by randomizing the field producing the SOC.

The combined effect of spin-independent disorder and random SOC on the localization in two-dimensional lattices has been 
studied in Refs. \cite{Evangelou1995,Asada2002} and the orbital effect of the magnetic field in these systems 
was addressed in Ref. \cite{Wang2015}.
In this paper we consider a continuous one-dimensional system with 
randomness solely in the SOC realization, and investigate the effect of 
the Zeeman field on the particle spins and localization. 
We show that similarly to the conventional random potentials, SOC can lead to  
localization, here strongly dependent on the Zeeman field. In particular,
we show that as the Zeeman splitting increases, the spin expectation values change strongly and, more importantly, 
the fraction of the localized states rapidly decreases. 
This offers the ability to \textit{localize or delocalize the states 
solely by acting at the particle spin}. The weakening of the SOC effects in sufficiently strong Zeeman 
fields is known in atomic physics 
as the Paschen-Back (alias nonlinear Zeeman) effect \cite{Paschen,Landau,Bransden}. Here we study the appearance 
of the Paschen-Back effect in a random macroscopic system, where, along with the spin dependence, 
the SOC effective weakening manifests itself as the delocalization of the states
under increasing Zeeman splitting.

This paper is organized as follows. In Sec. II we introduce the random SOC field and present its main 
characteristics. In Sec. III we describe a general picture of the macroscopic random Paschen-Back effect.
In Sec. IV this approach will be applied to the ground state of the system. Section V provides conclusions  and  
outlook for future research. Some details of calculations and additional information are given in the Appendices.

\section{Hamiltonian, random fields, and their correlators} We consider a 
system described by the following Hamiltonian with a spatially random SOC
$\alpha(x)$: 
\begin{equation}
H_{0}=\frac{k^{2}}{2}+\frac{1}{2}\left( \alpha (x)k+k\alpha (x)\right) \sigma
_{z}+\Delta \sigma_{x} ,
\label{H0}
\end{equation}%
where $k=-i\partial/\partial{x},$ $2\Delta$ is the Zeeman splitting, 
and $\sigma_{x,z}$ are the Pauli matrices \cite{3D}. We use units $\hbar\equiv 1$ and 
particle mass $\equiv 1$, so that from now on all the quantities are expressed in dimensionless units. Since our results will be based on the probability and the 
spin density distributions, they are independent of the Zeeman field direction, 
provided that it is orthogonal to the $z-$axis.

To emphasize the physical mechanism of the delocalization in the Paschen-Back effect,
we use unitary transformation \cite{Levitov} $H_{\rm tr}=S^{-1}H_{0}S$ 
with $S=\exp \left(-iA(x)\sigma_{z}\right),$ 
to reduce the Hamiltonian to the form
\begin{equation}
\label{H_transformed}
H_{\rm tr}=\frac{k^{2}}{2}-\frac{1}{2}\alpha^{2}(x)
+\Delta\left[\sigma_{x}\cos2A(x)-\sigma_{y}\sin2A(x)\right],
\end{equation}
where 
\begin{equation}
A(x)\equiv\int_{-L}^{x}\alpha (x^{\prime })dx^{\prime},
\end{equation}%
as we study a system of size $2L$, with $x\in[-L,L]$ \cite{lengthL}.

At $\Delta=0,$ the Hamiltonian (\ref{H_transformed}) 
describes two decoupled spin components in the random potential $-\alpha^{2}(x)/2$. 
The case of not random $\alpha(x)$ has been studied in Refs.
\cite{Sanchez2006,Valin,Cserti}. 
The Zeeman coupling in Eq. \eqref{H_transformed} 
gives rise to an effective magnetic field which has a constant amplitude $\Delta$ and randomly varying
direction: $\mathbf{m}(x)=\left(\cos 2A(x),-\sin 2A(x) ,0\right)$. 

\begin{figure}[t]
\begin{center}
\includegraphics[width=0.9\columnwidth]{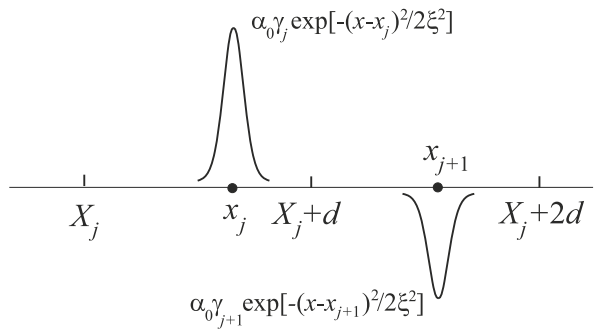}
\end{center}
\caption{Model distribution of SOC impurities corresponding
to the disorder in Eq. \eqref{alpha}.}
\label{fig:impurities}
\end{figure}

As a model of disorder we consider the following:
\begin{eqnarray}
\label{alpha}
\alpha (x)=\alpha_{0}\sum_{j=1}^{N}\gamma_{j}e^{-(x-x_{j})^{2}/2\xi^{2}}.
\end{eqnarray}
Here $j=1,\cdots,N$ labels $N$ ``impurities" having equal widths $\xi$ and located at points $x_{j}=X_{j}+d(r_{j}+1/2)$, where
$X_{j}\equiv\,-L+d(j-1)$,
the impurity concentration is $1/d,$ where $d=2L/N,$ 
and $\alpha_{0}\gamma_j$ are their strengths, as  shown in Fig. \ref{fig:impurities}. 
The statistics is described by independent uniform random distributions of
$\gamma_{j}$ and $r_{j}$, both in the range $[-0.5,0.5].$ 
The resulting potential is bounded, $\alpha^{2}(x)\le\alpha_{\rm max}^{2},$ and 
resembles optical speckles \cite{Falco}. 

At $L\gg\max\left(\xi,d\right),$ the random SOC is characterized 
by two main parameters. The first one is the mean square 
$\langle\langle\alpha^{2}\rangle\rangle,$ where $\langle\langle\ldots\rangle\rangle$ 
stands for the statistical averaging with the distributions of $r_{j}$ 
and $\gamma_{j}$. The second parameter is the correlation length $l_{\alpha}$ of $\alpha(x)$. 
In the model of Eq. \eqref{alpha}, 
$\langle\langle\alpha^{2}\rangle\rangle=\sqrt{\pi}\alpha_{0}^{2}\xi/12\,d$ and 
$l_{\alpha}=\sqrt{\pi}\xi$, as can be proven by straightforward calculations (e.g., by integration of the 
corresponding range function \cite{Glazov2011,Bindel2016}).

To describe the spatial scale of the random Zeeman field, we introduce the correlator 
${\cal K}_{mm}(x,x')\equiv\langle\langle \mathbf{m}(x)\mathbf{m}(x')\rangle\rangle$
and define the characteristic 
length $l_{\mathbf{m}}$ on which the direction
$\mathbf{m}(x)$ varies significantly. This is the distance at which the correlation 
between $\mathbf{m}(x)$ and $\mathbf{m}(x+l_{\mathbf{m}})$ becomes weak, that is 
$|{\cal K}_{mm}(x,x')|\ll 1$ for $|x-x'|\gg\,l_{\mathbf {m}}.$ 
The respective calculations can be done by taking into account that $A(x)$ is a random walk ~\cite{Norris}
in the $(x,A(x))-$space with small uncorrelated ``steps'' of 
the order of $\langle\langle \alpha^{2}\rangle\rangle^{1/2}l_{\alpha}\ll 1$ 
at the length scale of $l_{\alpha}.$ In this way we obtain (see Appendix A) 
\begin{equation}
l_{\mathbf{m}}=\frac{1}{4\left\langle\langle\alpha^{2}\right\rangle\rangle l_{\alpha }}.
\label{lm}
\end{equation}%


\section{Zeeman field-dependence: a general picture} For $\Delta\neq\,0$ 
the eigenstates of the 
Hamiltonian \eqref{H0} are nondegenerate (except for accidental events). 
Such states are characterized by spinors ${\bm\psi}_{n}(x)=\left[\psi_{n1}(x),\psi_{n2}(x)\right]^{\mathrm{T}},$
where the number $n=0,1,\cdots$ labels their energies $E_{n}.$
The spatial extension of state $n$ is characterized by the 
inverse participation ratio (IPR) \cite{IPR}: 
\begin{equation}
\zeta_{n}=
\int_{-L}^{L}\left[\left|\psi_{n1}(x)\right|^2+\left|\psi_{n2}(x)\right|^2\right]^{2}dx.
\label{ipr}
\end{equation}%
The symmetry of the Hamiltonian \eqref{H0} implies that the 
only nonzero mean spin component is given by%
\begin{equation}
\langle\sigma_{x}\rangle_{n}
=2\mathrm{Re}\int_{-L}^{L}\,\psi_{n1}^{\ast }(x)\psi_{n2}(x)dx.
\label{sigmax}
\end{equation}%
Note that the eigenfunctions of \eqref{H0} and \eqref{H_transformed} 
are mixed states in the spin subspace resulting 
in $\langle\sigma_{x}\rangle_{n}^{2}\le\,1$ with $\langle\sigma_{x}\rangle_{n}^{2}=1$
for a pure and $\langle\sigma_{x}\rangle_{n}^{2}=0$ for the maximally mixed state, 
respectively.

\begin{figure}[t]
\begin{center}
\includegraphics[width=0.9\columnwidth]{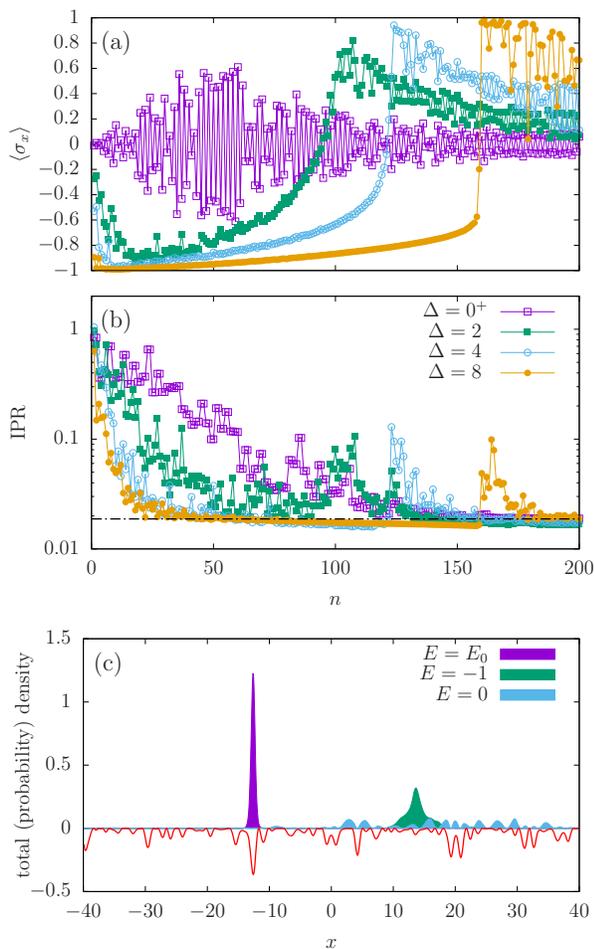}
\end{center}
\caption{(a) Spin component $\langle\sigma_{x}\rangle_{n}$ and (b) log-scale of the IPR as
a function of the state number for different Zeeman fields (the legend is shown in (b)). To avoid degeneracy,
we use here $0^{+}=10^{-3}.$ The horizontal dash-dotted line corresponds to $\zeta_{L}=3/4L$ 
value. 
(c) Actual realization of the random potential (solid line) and three densities
corresponding to the energies $E_{n}=E_{0},-1,$ and $0$. 
Here $\alpha_{0}=4$, $d=\xi=0.5,$ and $L = 40$.
}
\label{fig:single}
\end{figure}

Figure \ref{fig:single} presents the spin (a) and the IPR (b) as a function of $n,$ 
for a single realization of the random potential, which is shown in Fig. \ref{fig:single}(c). 
In Fig.~\ref{fig:single} (a) we observe that at small $\Delta$ most of the 
states are strongly mixed
in spin subspace with $\left|\langle\sigma_x\rangle_{n}\right|\ll\,1.$ 
By increasing $\Delta$, high-purity states appear at energies close to $\pm\Delta$ with 
$\langle\sigma_{x}\rangle_{n}$ increasing from approximately $-1$ to $1$ with the energy increase
from $-\Delta$ to $\Delta$. 
The IPR of well-localized states, namely those with $\zeta_{n}\gtrsim 0.7,$ strongly
varies as a function of the state number \cite{Evers} and 
reaches the disorder-free value $\zeta_{L}=3/4L$ at sufficiently large $n.$
For nonzero $\Delta,$ the $n$-dependence of the IPR becomes more
narrow, corresponding to the delocalization. 

\begin{figure}[t]
\begin{center}
\includegraphics[width=0.9\columnwidth]{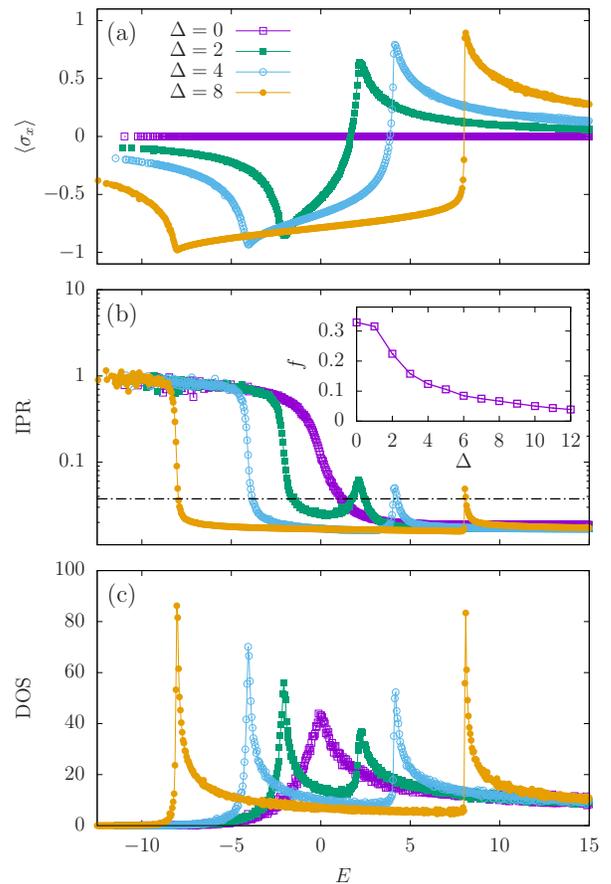}
\end{center}
\caption{Disorder-averaged quantities as a function 
of the state energy for different Zeeman $\Delta'$s (the legend is shown in (a)). 
(a) Expectation value $\langle\sigma_{x}\rangle_{n},$ (b) the IPR, where the 
inset shows the fraction $f$ of localized states (out of 300 lowest eigenstates) 
with the $\zeta_{n}>2\zeta_{L}$ (dash-dot horizontal line), and (c)
the density of states. The averaging is
performed over $10^{3}$ $\alpha(x)$ realizations with the parameters same 
as in Fig. \ref{fig:single}. }
\label{fig:averaged}
\end{figure}

Figure \ref{fig:averaged} shows the disorder-averaged spin (a),
the IPR (b), and the density of states (c) as a function of the energy. 
The IPR shows an \textit{effective} mobility edge \cite{Sanchez}, 
which sharpens and shifts approximately to $-\Delta$ as $\Delta$ increases. 
As shown in the panel (c), at $\Delta =0$ one observes a strong low-energy tail
in the density of localized states. By increasing $\Delta,$ the number of the 
states in the tail decreases, demonstrating the delocalization, as clearly seen also 
in the inset of the panel (b). 

To understand qualitatively the effect of the Zeeman field on delocalization, 
let us denote by $l_{s}$ the distance that a particle can travel under the influence of 
the random magnetic field before its spin becomes uncorrelated with the initial one.
By using again the random-walk approach, now in the coordinate-spin space,
for a semiclassical particle moving with the velocity $v,$ we obtain
that $l_{s}$ is determined by the condition 
$\Delta^{2}\left(l_{\mathbf m}/v\right)^{2}l_{s}/l_{\mathbf{m}}\sim 1,$ 
so that it is natural to define 
\begin{equation}
l_{s}\equiv \frac{v^{2}}{\Delta^{2}l_{\mathbf{m}}}=
4\frac{v^{2}\langle\langle\alpha^{2}\rangle\rangle l_{\alpha }}{\Delta^{2}},
\label{lS}
\end{equation}%
where $l_{\mathbf{m}}$ is given by Eq. \eqref{lm}. For states with energies $E_{n}$ 
close to zero such that $|E_{n}|\ll \langle\langle
\alpha^{2}\rangle\rangle,$ we can make a semiclassical estimate 
$v^{2}\sim\langle\langle \alpha^{2}\rangle\rangle $ and obtain
$l_{s}\sim {\langle\langle \alpha^{2}\rangle\rangle^{2}l_{\alpha}}/{\Delta^{2}}$.

Because long-range localization with $\zeta_{n}\ll\,l_{\alpha}$ 
occurs as a result of interference of waves 
with the same spin scattered by disorder \cite{Anderson,Berezinskii}, these localized states should have 
the characteristic length $1/\zeta_{n}\lesssim l_{s}.$ Thus, the random Zeeman field can destroy 
the localization \cite{Hikami}. On qualitative level, the destructive effect of decrease in $l_{s}$ with $\Delta$ 
is seen in Figs. \ref{fig:single}(b) and \ref{fig:averaged}(b).
Here, the states with $\zeta_{n}\gtrsim l_{s}^{-1}$
are still localized, while the higher-energy states are already delocalized, leading to the 
observed sharpening of the 
effective mobility edge and shifting it to lower energies. 

Since Hamiltonian (\ref{H0}) depends on spin randomly, 
in addition to the above argument based on comparison of the scales of $\zeta_{n}^{-1}$ and $l_{s}$, 
the delocalization and the dependence of $\langle\sigma_{x}\rangle_{n}$ 
on $\Delta$ can be obtained as follows. Let us consider
the matrix form ${\cal H}_{pq}$ of Hamiltonian \eqref{H0} in the representation of the degenerate basis states at $\Delta=0$
defined as
\begin{equation}
\widetilde{\bm\psi}_{2m}\equiv
\left[
\begin{array}{c}
\phi_{m}(x) \\
0
\end{array}
\right]
e^{-iA(x)}, \quad 
\widetilde{\bm\psi}_{2m+1}\equiv
\left[
\begin{array}{c}
0 \\
\phi_{m}(x) 
\end{array}
\right]
e^{iA(x)},
\label{basis}
\end{equation}%
where $\phi_{m}(x)$ ($m=0,1,\ldots$) are the real eigenfunctions 
with $\phi_{m}^{\prime\prime}(x)=-\left(\alpha^{2}(x)+2\epsilon_{m}\right)\phi_{m}(x),$ and 
eigenenergies $\epsilon_{m}.$ In this basis the diagonal 
components are: ${\cal H}_{2m,2m}={\cal H}_{2m+1,2m+1}=\epsilon_{m}$
and the off-diagonal ones are expressed as: 
\begin{equation}
{\cal H}_{2m+1,2l}^{*}={\cal H}_{2m,2l+1}\equiv\Delta\int_{-L}^{L}\phi_{m}(x)\phi_{l}(x)
e^{2iA(x)} dx.
\label{Hml}
\end{equation}
A broad Fourier spectrum of random $A(x)$ leads to appreciable transition 
coefficients ${\cal H}_{pq}/\Delta$ for localized states, which would
be negligibly small otherwise even if such states 
have a considerable spatial overlap. 
This possibility of particle transfer between different states
leads to delocalization at sufficiently strong $\Delta.$

\begin{figure}[t]
\begin{center}
\includegraphics[width=0.7\columnwidth]{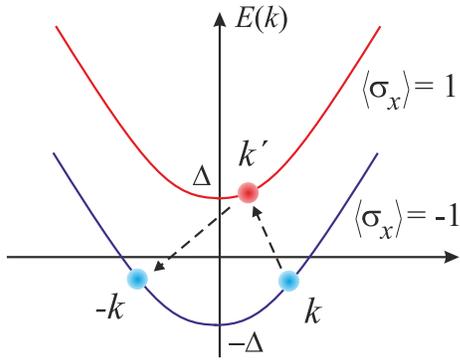}
\end{center}
\caption{Schematic illustration of the spin-conserving backscattering 
caused by the random SOC. Lower and upper parabolas correspond to $k^{2}/2-\Delta$ and $k^{2}/2+\Delta$
branches, respectively, with the virtual transitions shown by dashed lines.}
\label{fig:scattering}
\end{figure}

Now we can consider strong Zeeman field in more detail by addressing 
the source of suppression of the spin-conserving backscattering with the increase in $\Delta$.
At sufficiently large $\Delta$, 
neglecting the SOC, the single particle states 
can be presented as $\left|k,\langle\sigma_{x}\rangle\right\rangle,$ 
with $\langle\sigma_{x}\rangle=\pm\,1,$ corresponding to the eigenstates of $\sigma_{x}$ in Eq. \eqref{H0}, 
momentum $k,$ and energy $k^{2}/2+\langle\sigma_{x}\rangle\Delta.$
We consider the random SOC as a perturbation, which, however, 
prohibits the spin-conserving backscattering as the first-order process. 
Here this scattering $|k,-1\rangle\rightarrow|-k,-1\rangle$
occurs only by involving intermediate $|k^{\prime},1\rangle$ states
with the opposite spin, as schematically illustrated in Fig. \ref{fig:scattering}. 
The corresponding spin-conserving 
backscattering matrix element behaves for $k^{2}\ll 4\Delta$ 
as $\sim 1/\Delta,$ strongly decreasing the scattering 
probability for low-energy states (see Appendix B)  with the increase in $\Delta$ and, thus, leading to 
the delocalization.

\section{Ground state dependence on the Zeeman field} Now we consider how the 
developed approach can be applied to the properties of the ground state.
According to the Hellmann-Feynman theorem \cite{Feynman}, 
the expectation value of the spin of the ground state can be written
as: $\langle\sigma_{x}(\Delta=0)\rangle_{0}=(\partial E_{0}/\partial\Delta)_{\Delta=0}$, and, therefore 
obtained by the $\Delta-$perturbation theory for the ground state energy. 

\begin{figure}
\begin{center}
\includegraphics[width=0.9\columnwidth]{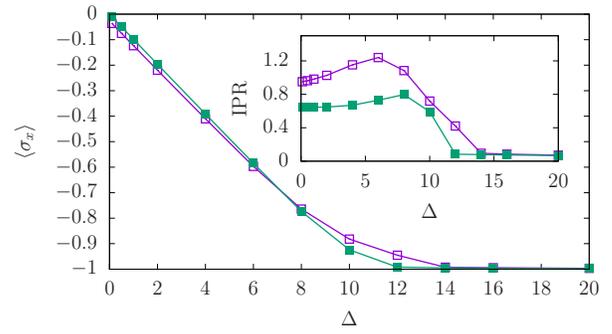}
\end{center}
\caption{Dependence of the ground-state $\langle\sigma_{x}(\Delta)\rangle_{0}$ (main plot)
and the IPR (inset) for two typical realizations of the random potential (see Appendix C for more details). 
As expected for low purity spin states, 
$\left|\langle\sigma_{x}\left(0\right)\rangle_{0}\right|\ll 1,$ corresponding to typical $l_{0}|\alpha(x_{0})|\ge 1$ 
for the chosen parameters of disorder, here the same as in Fig. \ref{fig:single}.}
\label{fig:ground}
\end{figure}

We begin by assuming that the Zeeman field is sufficiently weak such that 
the ground state spin can be written as: $\langle\sigma_{x}\left(\Delta\right)\rangle_{0}=
\langle\sigma_{x}\left(0\right)\rangle_{0}+
\Delta\,\partial\langle\sigma_{x}\left(\Delta\right)\rangle_{0}/\partial\Delta,$ where the derivative is calculated at $\Delta=0$.
Here the spin-split ground state forms a doublet well-separated from the rest of the states.
By using perturbation theory for degenerate states \cite{Landau} in the basis of Eq. \eqref{basis} 
we obtain the ground state: 
\begin{equation}
{\bm\psi}_{0}(x)=\frac{1}{\sqrt{2}}\phi_{0}(x)
\left[ 
\begin{array}{c}
\exp\left[-i(A(x)-\chi_{0}/2)\right] \\ 
-\exp \left[i(A(x)-\chi_{0}/2)\right]
\end{array}%
\right],
\label{psind}
\end{equation}%
where the phase $\chi_{0}$ is defined by 
${\cal H}_{01}\equiv\left\vert {\cal H}_{01}\right\vert\exp\left(i\chi_{0}\right).$ 
The condition of this weak-field approximation is $
\max\left(|{\cal H}_{0,2m+1}|/(\epsilon_{m}-\epsilon_{0})\right)\ll 1$ 
for $m\ge 1.$

To find $\langle\sigma_{x}\left(0\right)\rangle_{0},$ we assume that 
the ground state wave function is localized near a point $x_{0}$ and can be
approximated by a Gaussian of width $l_{0}$ as: 
$\phi_{0}^{2}(x)\approx\exp\left[-\left(x-x_{0}\right)^2/l_{0}^{2}\right]/\pi^{1/2}l_{0}.$ 
Next, by using ${\bm\psi}_{0}(x)$ in Eq. \eqref{psind} and approximating 
$A(x)\approx\,A(x_{0})+\alpha\left(x_{0}\right) \left( x-x_{0}\right)$ 
we obtain by Eq. \eqref{sigmax}:
\begin{equation}
\langle \sigma_{x}\left( 0\right) \rangle_{0}=-\exp\left(
-\alpha^{2}\left(x_{0}\right)l_{0}^{2}\right).
\end{equation}%
This value, being exponentially dependent on the
ground state parameters, strongly varies from realization 
to realization (see Fig. \ref{fig:ground}). 
To get an order-of-magnitude estimate of $\langle \sigma_{x}\left( 0\right) \rangle_{0}$ 
we consider a model ground state in the 
potential characterized by $\gamma_{j}=r_{j}=0.5$ and $\gamma
_{j+1}=-r_{j+1}=0.5.$ This state has the 
width $l_{0}=\sqrt{{\xi }/{\alpha_{0}}}$ yielding 
$\langle \sigma_{x}\left( 0\right) \rangle_{0}=-\exp \left( -\alpha
_{0}\xi \right)$. Next, we calculate the inverse participation ratio for this 
state as: 
\begin{equation}
\zeta_{0}\left( 0\right) =\sqrt{\frac{\alpha_{0}}{2\pi \xi }}.
\end{equation}%
For given system parameters this yields $\zeta_{0}(0)\approx 1.12,$ similar to
the numerical results in the inset of Fig. \ref{fig:ground}.

Next, by means of the second-order perturbation theory 
and the Hellmann-Feynman theorem, one can obtain the linear in $\Delta$ term in 
$\langle\sigma_{x}\left(\Delta\right)\rangle_{0}.$ To this end, we 
calculate $\Delta^{2}-$correction to the 
energy by summing up over all transitions to the higher-energy states in the Eq. \eqref{basis} basis. 
The maximal contribution to the energy correction is achieved at the states 
with energies $2\alpha^{2}(x_{0}),$ lying high above the effective mobility edge. 
Such states can be accurately approximated 
as $\sin(kx+\delta)/\sqrt{L}$, extended to the total length of the system with a slowly varying 
phase $\delta.$ 
The energy calculation can be done analytically by using the steepest descent method \cite{Mathews}
(provided that $2\alpha(x_{0})\,l_{0}\gg\,1$) resulting in 
\begin{equation}
\frac{d\langle\sigma_{x}\left(\Delta\right)\rangle_{0}}{d\Delta}=
-\frac{2}{\left\vert \epsilon_{0}\right\vert +2\alpha^{2}\left( x_{0}\right)}.
\end{equation}%
This value is less sensitive to the disorder realization 
than $\langle\sigma_{x}(0)\rangle_{0},$ as can be seen from the slope 
of $\langle\sigma_{x}\left(\Delta\right)\rangle_{0}$ in Fig.~\ref{fig:ground}, presenting 
the numerical evidence for the random Paschen-Back effect. 
As it is seen in the main plot, 
$\langle\sigma_{x}(\Delta)\rangle_{0}$ 
tends to $-1$ at sufficiently large $\Delta,$ as expected for the 
conventional Paschen-Back effect \cite{Paschen}. 
Note that even at rather small $\Delta$, the linear term greatly exceeds $\langle\sigma_{x}(0)\rangle_{0}.$
The IPR shown in the inset initially increases (see Appendix), corresponding 
to a stronger localization, and then decreases to the values $\sim\zeta_{L}$, 
demonstrating the delocalization.

\section{Conclusions and outlook} We have studied the dependence of
single-particle states on the Zeeman field in a one-dimensional system with
random spin-orbit coupling. The observed dependence of the spin is nonlinear 
with the saturation at a sufficiently strong field, corresponding to 
a macroscopic random Paschen-Back effect. In such a system, the 
spin saturation is accompanied by particle delocalization as both 
effects are due to suppression of the role of the random spin-orbit 
coupling. These effect could be engineered in a broad range of parameters in experimental setups 
for cold atomic gases, therefore permitting a variety of studies of this fundamental quantum effect at 
a macroscopic level. Although the calculated quantities are based on a particular model of disorder, 
our main estimates and qualitative results, being obtained by means of general arguments, are not restricted to the chosen model.

\begin{acknowledgments}
M.M. and E.Y.S. acknowledge the support by the Grant FIS2015-67161-P (MINECO of Spain/FEDER) and 
Grupos Consolidados UPV/EHU del Gobierno Vasco (IT-986-16). 
V.V.K. acknowledges the support of the FCT (Portugal) under the grant UID/FIS/00618/2013.
E.Y.S. is grateful to V.K. Dugaev and M.M. Glazov for valuable discussions.
\end{acknowledgments}

\appendix

\section{Correlator of the random magnetic field}

We present the correlator of the directions of the random magnetic field $%
\mathcal{K}_{mm}(x^{\prime},x)\equiv \langle \langle \mathbf{m}(x^{\prime})\mathbf{m}(x)\rangle\rangle$ 
as 
\begin{align}
\mathcal{K}_{mm}(x^{\prime },x)&=\langle \langle 
\cos\left[2\left(A(x^{\prime})-A(x)\right) \right]\rangle\rangle 
\nonumber\\
&=\mathrm{Re}\left\langle
\left\langle \prod_{j}\exp \left[ 2i\int_{X_{j}}^{X_{j}+d}\alpha (y)dy\right]
\right\rangle \right\rangle,
\label{Kmm}
\end{align}%
using the product over single-impurity intervals $\left(
X_{j},X_{j}+d\right) $ (as shown in Fig. \ref{fig:impurities}), located between points $x^{\prime}$ 
and $x$ and note that the distribution in Eq. \eqref{alpha} allows one to separate calculations
of products and averaging. Taking a single interval and assuming for
simplicity $\xi \ll d$ with 
\begin{equation}
J_{j}\equiv 2\int_{X_{j}}^{X_{j}+d}\alpha (y)dy=2\sqrt{2\pi }\gamma
_{j}\alpha_{0}\xi ,  \label{Jj}
\end{equation}%
yields 
\begin{equation}
e^{iJ_{j}}=\cos \left( 2\sqrt{2\pi }\gamma_{j}\alpha_{0}\xi \right) +i\sin
\left( 2\sqrt{2\pi }\gamma_{j}\alpha_{0}\xi \right) .
\end{equation}%

\begin{figure}
\begin{center}
\includegraphics[width=0.9\columnwidth]{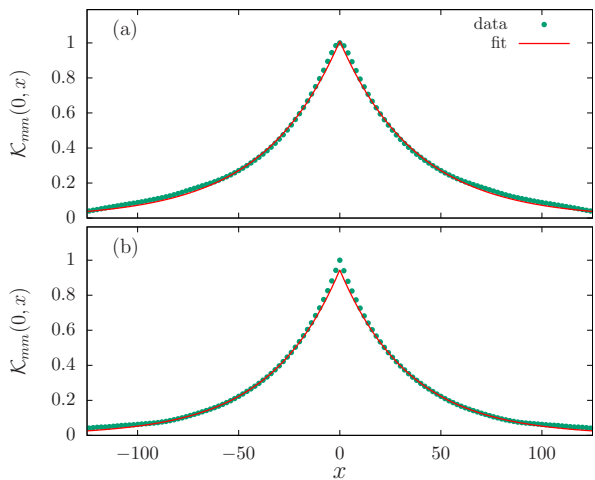} 
\end{center}
\caption{Correlator $\mathcal{K}_{mm}(0,x)$ (averaged over $10^{3}$ realizations) 
for two random potentials, both
with $d=0.5$. (a) $\protect\alpha_{0}=1,\,\protect\xi =d/4$ and best
fitting parameter $\protect\beta =0.03$, (b) $\protect\alpha_{0}=0.06125,%
\protect\xi =4d$ and best fitting parameter $\protect\beta =0.026.$}
\label{fig:correlator}
\end{figure}

Since in the model of disorder we are considering, the expectation value $%
\left\langle \gamma_{j}\right\rangle =0,$ one obtains $\left\langle
\left\langle \sin (2\sqrt{2\pi }\gamma_{j}\alpha_{0}\xi )\right\rangle
\right\rangle =0.$ Employing a ``small change'' approximation $\alpha_{0}\xi \ll 1$ we obtain 
\begin{equation}
\left\langle \left\langle \cos (2\sqrt{2\pi }\gamma_{j}\alpha_{0}\xi
)\right\rangle \right\rangle =1-4\pi \left\langle \gamma
_{j}^{2}\right\rangle \left( \alpha_{0}\xi \right)^{2}
+{\cal O}\left((\alpha_{0}\xi)^{4}\right).
\end{equation}%
Making $\gamma_{j}-$averaging with $\left\langle \gamma
_{j}^{2}\right\rangle =1/12$ and taking into account that $\left\langle
\left\langle \alpha^{2}\right\rangle \right\rangle =\sqrt{\pi }/12\times
\alpha_{0}^{2}\xi /d$ yields with the same accuracy:
\begin{equation}
\left\langle \left\langle \cos (2\sqrt{2\pi }\gamma_{j}\alpha_{0}\xi
)\right\rangle \right\rangle =1-4\sqrt{\pi }\left\langle \left\langle \alpha
^{2}\right\rangle \right\rangle \xi d.
\end{equation}%
Next, we build the product over the intervals and obtain for $x^{\prime }=0$ and $%
d\ll \left\vert x\right\vert \ll L$ ($2L$ is the total system length): 
\begin{equation}
\mathcal{K}_{mm}(0,x)=\left( 1-4\sqrt{\pi }\left\langle \left\langle \alpha
^{2}\right\rangle \right\rangle \xi d\right)^{\left\vert x\right\vert
/d}\approx \exp \left( -\beta \left\vert x\right\vert \right) ,
\end{equation}%
where $\beta =4\sqrt{\pi }\left\langle \left\langle \alpha^{2}\right\rangle
\right\rangle \xi .$ The corresponding correlation length can be defined as:%
\begin{equation}
l_{\mathbf{m}}=\int_{0}^{\infty }\mathcal{K}_{mm}(0,x)dx=\frac{1}{4\sqrt{\pi 
}\left\langle \left\langle \alpha^{2}\right\rangle \right\rangle \xi },
\label{lmS}
\end{equation}%
where we put the upper integration limit to infinity and the lower limit to
zero since we assume that $\l_{\alpha }\ll l_{\mathbf{m}}\ll L.$ By noting
that in our model of disorder the correlation length of the spin-orbit
coupling $l_{\alpha }=\sqrt{\pi }\xi $, we arrive at Eq. \eqref{lm}.
While the coefficient $4\sqrt{\pi }$ in Eq. \eqref{lmS} depends on the
details of the model of disorder, the $l_{\mathbf{m}}\sim 1/\left\langle
\left\langle \alpha^{2}\right\rangle \right\rangle \xi $ scaling is
model-independent. The numerical results are presented in Fig. \ref%
{fig:correlator} for two different sets of parameters. Note that at these
values of $\alpha_{0},\xi,$ and $d$ one obtains $\beta \approx 0.033$ in
agreement with the best fit of $\mathcal{K}_{mm}(0,x)$ (see caption of Fig. \ref{fig:correlator}).

Having established the long-range behavior of the correlator, it would be of
interest to obtain its short-distance behavior at $\left\vert x-x^{\prime}\right\vert \ll l_{\alpha }$. 
Taking into account that at these short
distances $A(x)-A(x^{\prime})\approx \alpha (x)(x-x^{\prime }),$ we obtain
after averaging of $\langle\langle\cos\left[2\left(A(x^{\prime})-A(x)\right)\right]\rangle\rangle$ 
in Eq. \eqref{Kmm} 
\begin{equation}
\mathcal{K}_{mm}(x^{\prime },x)=1-2\left\langle \left\langle \alpha
^{2}\right\rangle \right\rangle (x-x^{\prime })^{2}.  \label{Kshort}
\end{equation}
Note that short- and long-range behavior of $\mathcal{K}_{mm}(x,x^{\prime })$
is due to different spatial scales. The long-range behavior is determined by 
$l_{\mathbf{m}}$ in Eq. \eqref{lmS} while the short-range one \eqref{Kshort}
is determined by the length $1/\left\langle \left\langle \alpha
^{2}\right\rangle \right\rangle^{1/2}.$ For the choice of parameters in
Fig. \ref{fig:correlator} we have $l_{\mathbf{m}}\gg 1/\left\langle
\left\langle \alpha^{2}\right\rangle \right\rangle^{1/2},$ leading to a
cusp-like dependence presented in this Figure.

\section{Spin-conserving backscattering matrix element: spin-orbit coupling
as a perturbation}

Here we illustrate the $\Delta-$dependence of the spin-conserving
backscattering in the random spin-orbit coupling field and demonstrate that
its probability rapidly decreases with the increase in $\Delta$. We assume
strong Zeeman field limit, which determines the spin states and the
scattering due to the random spin-orbit couping.

We consider spin-conserving transition $\left\vert k,\sigma
_{x}=-1\right\rangle \rightarrow \left\vert -k,\sigma_{x}=-1\right\rangle,$
which occurs at $k^{2}<4\Delta $ via virtual transitions to intermediate $%
\left\vert k^{\prime},\sigma_{x}=1\right\rangle$ states, as shown in Fig. \ref%
{fig:scattering}. Using second-order perturbation theory we obtain for the
spin-conserving backscattering matrix element $M_{k}$ resulting from
interactions with random spin-orbit coupling impurities:

\begin{equation}
M_{k}=\frac{1}{4}\int_{-\infty }^{\infty }\alpha_{q}\alpha_{2k+q}\frac{%
\left( 2k+q\right)q}{2\Delta +\left( k+q\right)^{2}/2-k^{2}/2} \,\frac{dq}{2\pi},
\label{Mk}
\end{equation}%
where $q=k^{\prime}-k,$ and we have taken into account that the single spin-flip scattering matrix
element between $k$ and $k^{\prime }$ states is equal to $\alpha_{k^{\prime }-k}\left(k+k^{\prime}\right)2$ 
\cite{Dugaev09}, with the Fourier-component 
\begin{equation}
\alpha_{p}\equiv\int_{-\infty }^{\infty }\alpha (x)e^{-ipx}dx.
\end{equation}

The impurities have a Gaussian shape with the amplitude $\left\vert \gamma_{j}\right\vert =1$
resulting in: $\alpha_{p}=\sqrt{2\pi }\alpha_{0}\xi e^{-p^{2}/2\xi^{2}}$
with $\alpha_{q}\alpha_{2k+q}=2\pi \alpha_{0}^{2}\xi^{2}e^{-\left(
q^{2}+\left( 2k+q\right)^{2}\right) \xi^{2}/2}.$ Assuming a sufficiently large width $\xi$
such that $\exp \left( -\Delta \xi^{2}\right) \ll 1,$
we can use the steepest descent method to calculate the integral in Eq. \eqref{Mk}, where the
maximum backscattering probability is due the "symmetric" transition with
the momentum of the intermediate state $k+q=0.$ As a result, 
we obtain for the matrix element for the states near the bottom of the 
$-\Delta $ subband 
\begin{equation}
M_{k}=-\frac{\sqrt{\pi }}{8}\alpha_{0}^{2}\frac{\xi }{\Delta }%
k^{2}e^{-k^{2}\xi^{2}}.
\end{equation}%
This value of $\left\vert M_{k}\right\vert^{2}$ rapidly decreases with the
increase in $\Delta $ leading to delocalization by the Zeeman field.

\section{$\Delta -$dependence of the inverse participation ratio}

We begin with the study of the $\Delta -$dependence of the ground state
inverse participation ratio (IPR) in the limit of weak Zeeman field, where
the analysis can be done perturbatively. We seek for the ground state $\widetilde{\bm\psi}_{0}(x)$ in the form: 
\begin{align}
\widetilde{\bm\psi}_{0}(x)&=\frac{\sqrt{1-\nu }}{\sqrt{2}}\left[ 
\begin{array}{c}
\psi_{0}(x)e^{i\chi_{0}/2} \\ 
-\psi_{0}^{\ast }(x)e^{-i\chi_{0}/2}%
\end{array}%
\right] 
\nonumber\\
&\quad+\frac{1}{\sqrt{2}}\sum_{k}\left[ 
\begin{array}{c}
p_{k}\psi_{k}(x)e^{i\chi_{k}/2} \\ 
-p_{k}^{\ast }\psi_{k}^{\ast }(x)e^{-i\chi_{k}/2}%
\end{array}%
\right], 
\end{align}%
where $\psi_{0}$ is the ground state wave function in the $\Delta =0$ limit
with the energy $\epsilon_{0}$ (cf. Eq.  \eqref{basis}) and the
functions $\psi_{k}(x)$ are extended over the system length $2L$ wave functions
of the quasi-continuous spectrum with $%
\epsilon_{k}=k^{2}/2$. Small coefficients $p_{k}$ can be obtained by
perturbation theory as: 
\begin{equation}
p_{k}=\frac{\Delta }{\epsilon_{k}-\epsilon_{0}}\eta_{k},
\end{equation}%
where 
\begin{equation}
\eta_{k}=e^{-i\left( \chi_{0}+\chi_{k}\right) /2}\int_{-L}^{L}\psi
_{0}^{\ast }(x)\psi_{k}^{\ast }(x)dx.
\end{equation}%
The parameter $\nu$ is a small probability to find the particle in a
delocalized state:
\begin{equation}
\nu =\sum_{k}\left\vert p_{k}^{2}\right\vert =\frac{L}{\pi }\int_{-\infty
}^{\infty }\left\vert p_{k}^{2}\right\vert dk,
\end{equation}%
to conserve the total norm of the wavefunction. The probability $\nu $ can
be calculated by the steepest descent method similarly to the second-order
correction to the ground state energy assuming the Gaussian ground state 
with the maximum probability density at $x_{0}$ point as:%
\begin{equation}
\nu =\frac{\Delta^{2}}{\left( \left\vert \epsilon
_{0}\right\vert +2\alpha^{2}\left( x_{0}\right) \right)^{2}}.
\end{equation}%
Function $\vert\widetilde{\bm\psi}_{0}(x)\vert^{4}$ has a complex
structure, with, however, only two terms giving finite contribution to the
IPR in the $L\rightarrow \infty $ limit, as can be seen by counting the
powers of $L$ in the corresponding terms. The relevant contributions can be
presented in the form:%
\begin{align}
\label{psi4}
\left\vert\widetilde{\bm\psi}_{0}(x)\right\vert^{4}&=\left( 1-\nu \right)^{2}\left[
\left\vert \psi_{0}(x)\right\vert^{4}+2\left\vert\psi_{0}(x)\right\vert
^{2}\times\right.
\\
&\quad\left.
\left(\psi_{0}^{\ast}(x)\psi_{k}(x)e^{-i\chi_{0}/2}e^{i\chi
_{k}/2}p_{k}^{\ast }+\rm{c.c.}\right)\right].
\nonumber
\end{align}%
Here we concentrate on these terms, having different orders in $\Delta $ and
present the inverse participation ratio in the form of the $\Delta -$%
expansion:%
\begin{equation}
\zeta_{0}(\Delta )=\zeta_{0}(0)+\zeta_{0}^{\prime }(0)\Delta +\frac{1}{2}%
\zeta_{0}^{\prime \prime }(0)\Delta^{2}.
\end{equation}%
By using Eq. \eqref{psi4}, the term quadratic in $\Delta $ can be
rewritten as:%
\begin{equation}
\frac{1}{2}\zeta_{0}^{\prime \prime }(0)\Delta^{2}=-2\zeta_{0}(0)\nu ,
\end{equation}%
leading to a decrease in $\zeta_{0}(\Delta )$ with the increase in the
Zeeman field, as expected in delocalization scenario.

\begin{figure}[h]
\begin{center}
\includegraphics[width=0.9\columnwidth]{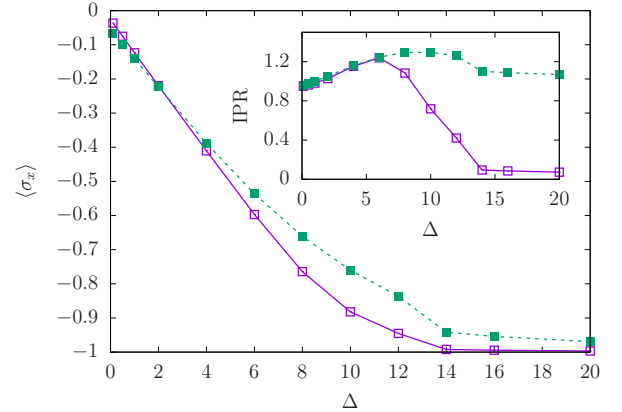}
\end{center}
\caption{Dependence of the ground state spin on the Zeeman $%
\Delta $ for random (solid line) and regular (as in Eq. \eqref{constalpha},  dashed line)  
SOC. These dependences are very similar for both types of coupling. Inset
shows the qualitative difference between the IPR for the random and the regular realizations. 
While at small $\Delta $ the behavior of
the IPR is the same, their large $\Delta-$dependences are different: the IPR
rapidly decreases for the random SOC and returns to its value at $\Delta=0$ for the regular
one.}
\label{fig:comparison}
\end{figure}

The term linear in $\Delta$ has the form: 
\begin{eqnarray}
&&\zeta_{0}^{\prime}(0)\Delta =2\times
\\
&&
\sum_{k}
\int_{-L}^{L}\left\vert\psi_{0}(x)\right\vert^{2}
\left(
\psi_{0}^{\ast}(x)\psi_{k}(x)e^{-i(\chi_{0}-\chi_{k})/2}p_{k}^{\ast} 
+\rm{c.c.}
\right) dx.
\nonumber
\end{eqnarray}
Note that while $\psi_{0}(x)$ and $\psi_{k}(x)$ are orthogonal, 
$\left|\psi_{0}(x)\right|^{2}\psi_{0}(x)$ and $\psi_{k}(x)$ are, in general, not. As a result we obtain
the linear correction to the IPR in the form:%
\begin{align}
\zeta_{0}^{\prime }(0)&=4\mathrm{Re}\sum_{k}\frac{e^{-i\chi_{0}}}{\epsilon
_{k}-\epsilon_{0}}\int_{-L}^{L}\psi_{0}^{\ast }(x)\psi_{k}^{\ast
}(x)dx
\times\\
&\quad
\int_{-L}^{L}\left\vert \psi_{0}(x)\right\vert^{2}\psi_{0}^{\ast
}(x)\psi_{k}(x)dx,
\nonumber
\end{align}%
demonstrating that IPR can behave linearly with $\Delta $, as presented in
Fig. \ref{fig:comparison}, due to change in the shape of the ground state
wave function by adding strongly $x-$dependent functions varying on the
spatial scale less than the spatial scale of $\psi_{0}(x)$.

One more point on the importance of disorder deserves to be mentioned here. To
demonstrate its role, we have chosen a realization of $\alpha (x)$ and
performed a calculation of the $\Delta -$dependent IPR of the ground state
with the Hamiltonian%
\begin{equation}
H=\frac{k^{2}}{2}+V(x)+\alpha (x_{0})k\sigma_{z}+\Delta \sigma_{z},
\label{constalpha}
\end{equation}%
where $V(x)=-\alpha^{2}(x)/2$ and $x_{0}$ is the position
of the maximum of the ground state density in this potential. Note that Hamiltonian \eqref{constalpha}  resembles
the Hamiltonian \eqref{H0}, but has a constant SOC. At sufficiently small $\Delta $ the properties of the ground state
are determined mostly by local SOC $\alpha(x_{0}).$ 
The effect of the randomness
becomes visible only at relatively large $\Delta,$ where the ground state
is already modified by a contribution of the extended states. Although in
both cases the value of spin saturates at $\langle\sigma_{x}\rangle=-1$, as expected in the conventional
Paschen-Back effect, the localization is restored
for a constant SOC and disappears for a random one, as can
be seen in Fig. \ref{fig:comparison}. This is due to different properties of
the interstate transition matrix elements (see Eq. \eqref{Hml}), 
where the broad Fourier spectrum of random $A(x)$ extends the set of transitions 
while for a regular coupling this set is strongly restricted and delocalization does not occur.

\begin{thebibliography}{99}

\bibitem{Dyakonov08} \textit{Spin Physics in Semiconductors} Springer Series
in Solid-State Sciences, Ed. by M. I. Dyakonov, Springer (2008).

\bibitem{spielman2013} T. D. Stanescu, B. Anderson, and V. Galitski, Phys.
Rev. A \textbf{78}, 023616 (2008); V. Galitski and I. B. Spielman, Nature \textbf{494},
49 (2013).

\bibitem{zhaih2012} H. Zhai, Int. J. Mod. Phys. B \textbf{26}, 1230001
(2012).

\bibitem{EXspielman2011} Y.-J. Lin, K. Jim\'{e}nez-Garc\'{\i}a, and I. B.
Spielman, Nature \textbf{471}, 83 (2011).

\bibitem{wang2012} P. Wang, Z.-Q. Yu, Z. Fu, J. Miao, L. Huang, S. Chai, H. Zhai, 
and J. Zhang, Phys. Rev. Lett. 109, 095301 (2012).

\bibitem{cheuk2012} L. W. Cheuk, A. T. Sommer, Z. Hadzibabic, T. Yefsah, 
W. S. Bakr, and M. W. Zwierlein, Phys. Rev. Lett. 109, 095302 (2012).

\bibitem{EXjin2012} J.-Y. Zhang, S.-C. Ji, Z. Chen, L. Zhang, Z.-D. Du, B.
Yan, G.-S. Pan, B. Zhao, Y.-J. Deng, H. Zhai, S. Chen, and J.-W. Pan, Phys.
Rev. Let. \textbf{109}, 115301 (2012).

\bibitem{EXqu2013} Ch. Qu, Ch. Hamner, M. Gong, Ch. Zhang, and P. Engels,
Phys. Rev. A \textbf{88}, 021604 (2013).

\bibitem{stringari2012} G. I. Martone, Y. Li, L. P. Pitaevskii, and S. Stringari,
Pys. Rev. A \textbf{86}, 063621 (2012).

\bibitem{zhang2012} Y. Zhang, L. Mao, and Ch. Zhang, Phys. Rev. Lett. 
\textbf{108}, 035302 (2012).

\bibitem{lu2013} Q.-Q. L\"{u} and D. E. Sheehy, Phys. Rev. A \textbf{88},
043645 (2013).

\bibitem{OL} J. Larson, J. P. Martikainen, A. Collin, and E. Sj\"oqvist, 
Phys. Rev. A, {\bf 82}, 043620;
Y. Zhang and C. Zhang, Phys. Rev. A \textbf{87}, 023611 (2013);
M. Salerno and F. Kh. Abdullaev, Phys. Lett. A {\bf 379},
2252 (2015); M. Salerno, F. Kh. Abdullaev, A. Gammal, and L. Tomio, Phys. Rev. A {\bf 94}, 043602 (2016) 

\bibitem{OL_ZL} Y. V. Kartashov, V. V. Konotop, D. A. Zezyulin, and L. Torner, 
Phys. Rev. Lett. \textbf{117}, 215301 (2016).

\bibitem{ZL} Y. V. Kartashov, V. V. Konotop, and F. K. Abdullaev, 
Phys. Rev. Lett. {\bf 111}, 060402 (2013); 
V. E. Lobanov, Y. V. Kartashov, and V. V. Konotop, 
Phys. Rev. Lett. {\bf 112}, 180403 (2014). 

\bibitem{LocSO} Y. V. Kartashov, V. V. Konotop, and D. A. Zezyulin, Phys. Rev. A {\bf 90}, 063621 (2014).

\bibitem{Anderson} P. W. Anderson, Phys. Rev. \textbf{109}, 1492 (1958).

\bibitem{Berezinskii} V. L. Berezinskii, 
Sov. Phys. JETP \textbf{38}, 620 (1974); L.P. Gor'kov, in: \emph{Electron-electron interactions in disordered systems}, 
pp. 619, Eds. A.L. Efros and M. Pollak, North-Holland, Amsterdam (1985).

\bibitem{Flach1} Ch. Skokos, D. O. Krimer, S. Komineas, and S. Flach, Phys.
Rev. E \textbf{79}, 056211 (2009).

\bibitem{Pikovsky} A. S. Pikovsky and D. L. Shepelyansky, Phys. Rev. Lett. 
\textbf{100}, 094101 (2008).

\bibitem{Larcher1} M. Larcher, F. Dalfovo, and M. Modugno, Phys. Rev. A 
\textbf{80}, 053606 (2009).

\bibitem{Aleiner} I. L. Aleiner, B. L. Altshuler, and G. V. Shlyapnikov,
Nat. Phys. \textbf{6}, 900 (2010).

\bibitem{Zhou} For one-dimensional systems see: L. Zhou, H. Pu, and W. Zhang, Phys. Rev. A \textbf{87},
023625 (2013) and for two-dimensional speckles: G. Orso, Phys. Rev. Lett. \textbf{118}, 105301 (2017).

\bibitem{LYKKC} C. Li, F. Ye, Y. V. Kartashov, V. V. Konotop, and X. Chen, 
Sc. Rep. {\bf 6}, 31700 (2016). 

\bibitem{mardonov2015} Sh. Mardonov, M. Modugno, and E. Ya. Sherman, 
 Phys. Rev. Lett. \textbf{115}, 180402 (2015).

\bibitem{Glazov2010} M. M. Glazov, E. Ya. Sherman, and V. K. Dugaev, Physica E \textbf{42}, 2157 (2010).

\bibitem{Glazov2011} M. M. Glazov and E. Ya. Sherman, Phys. Rev. Lett. \textbf{107}, 156602 (2011).

\bibitem{Bindel2016} J. R. Bindel, M. Pezzotta,	J. Ulrich, M. Liebmann,	E. Ya. Sherman, and M. Morgenstern,
Nat. Phys. \textbf{12}, 920 (2016).

\bibitem{Evangelou1995} S. N. Evangelou, Phys. Rev. Lett. \textbf{75}, 2550 (1995).

\bibitem{Asada2002} Y. Asada, K. Slevin, and T. Ohtsuki, Phys. Rev. Lett. \textbf{89}, 256601 (2002).

\bibitem{Wang2015} C. Wang, Y. Su, Y. Avishai, Y. Meir, and X. R. Wang, Phys. Rev. Lett. \textbf{114}, 096803 (2015). 

\bibitem{Paschen} F. Paschen and E. Back, Ann. der Physik \textbf{39}, 897 (1912); 
F. Paschen and E. Back, Ann. der Physik \textbf{40}, 960 (1913). 

\bibitem{Landau} L. D. Landau and E. M. Lifshitz, \textit{Quantum Mechanics} Butterworth-Heinemann Publ., 
Oxford, UK (1981) 

\bibitem{Bransden} B.H. Bransden and C.J. Joachain \textit{Physics of atoms and molecules} Longman Publ.,
New York, USA (1982). 

\bibitem{3D} Note that although the spatial motion is strictly one-dimensional, spin-related features are fully three-dimensional. 

\bibitem{Levitov} L. S. Levitov and E. I. Rashba, Phys. Rev. B \textbf{67}, 115324 (2003).

\bibitem{lengthL} The size of the system should be sufficiently larger than the disorder-induced 
localization length in the energy interval of interest. A practical check for a length choice is based on 
comparing systems of size $L$, $2L$, and $4L$, verifying that finite-$L$ effects and boundary conditions 
are not substantial.

\bibitem{Sanchez2006} D. S\'{a}nchez and L. Serra, Phys. Rev. B \textbf{74}, 153313 (2006);
D. S\'{a}nchez, L. Serra, and M.-S. Choi, Phys. Rev. B \textbf{77}, 035315 (2008).

\bibitem{Valin} M. Val\'{i}n-Rodriguez, A. Puente, and L. Serra, Phys. Rev. B \textbf{69}, 085306 (2004).

\bibitem{Cserti} J. Cserti, A. Csord\'{a}s, and U. Z\"{u}licke, Phys. Rev. B \textbf{70}, 233307 (2004).

\bibitem{Falco} G. M. Falco, A. A. Fedorenko, J. Giacomelli, and M. Modugno,
Phys. Rev. A \textbf{82}, 053405 (2010).

\bibitem{Norris} J. R. Norris \textit{Markov chains} 
(Cambridge series on statistical and probabilistic mathematics) Cambridge University Press (1998).

\bibitem{IPR} In contrast to lattice models \cite{Larcher1,Evangelou1995,Asada2002,Wang2015},
in a continous one-dimensional system, the IPR has units of inverse length being not limited by 1 from above. 
The IPR of the order of $1/2L$ corresponds to delocalized states in terms of the present model.

\bibitem{Evers} F. Evers and A. D. Mirlin, Phys. Rev. Lett. \textbf{84}, 3690 (2000).

\bibitem{Sanchez} The occurrence of an \textit{effective} mobility edge is typical of speckle-like potentials, see e.g.
L. Sanchez-Palencia, D. Cl\'{e}ment, P. Lugan, P. Bouyer, G. V. Shlyapnikov, and A. Aspect, 
Phys. Rev. Lett. \textbf{98}, 210401 (2007). For special shapes of correlated potentials 
demonstrating the mobility edge see: F. M. Izrailev and A. A. Krokhin, Phys. Rev. Lett. \textbf{82}, 4062 (1999).

\bibitem{Hikami} S. Hikami, A.I. Larkin, and Y. Nagaoka, 
Progr. of Theor. Phys. \textbf{63}, 707 (1980).

\bibitem{Feynman} R. P. Feynman, Phys. Rev. \textbf{56}, 340 (1939). 

\bibitem{Mathews} J. Mathews and R.L. Walker \textit{Mathematical methods of physics} W.A. Benjamin Inc., New York-Amsredam (1964). 

\bibitem{Dugaev09} V. K. Dugaev, E. Ya. Sherman, V. I. Ivanov, and J. Barna\'{s}, Phys. Rev. B \textbf{80}, 081301 (2009).

\end{thebibliography}
\end{document}